\documentclass[10pt,twocolumn,superscriptaddress,english,prl,showpacs,
floatfix,aps,nofootinbib]{revtex4-1}
\usepackage[utf8]{inputenc}
\usepackage{amsmath}
\usepackage{amssymb}
\usepackage{graphicx}
\usepackage{xspace}
\usepackage{mathtools}
\usepackage{comment}
\usepackage{physics}
\usepackage{bm}

\makeatletter
 
\usepackage[normalem]{ulem}
\usepackage{xcolor}
\usepackage{soul}
\usepackage[backref=none,
bookmarksnumbered=true,
bookmarks=true,
bookmarksopen=true,
colorlinks=true,
citecolor=blue,
linkcolor=blue,
anchorcolor=green,
urlcolor=blue,unicode=false]{hyperref}

\makeatother

\usepackage{babel}

\begin{document}

\title{Diode effects in current-biased Josephson junctions}

\author{Jacob F.\ Steiner}
\affiliation{\mbox{Dahlem Center for Complex Quantum Systems and Fachbereich Physik, Freie Universit\"at Berlin, 14195 Berlin, Germany}}

\author{Larissa Melischek}
\affiliation{\mbox{Dahlem Center for Complex Quantum Systems and Fachbereich Physik, Freie Universit\"at Berlin, 14195 Berlin, Germany}}

\author{Martina Trahms}
\affiliation{\mbox{Fachbereich Physik, Freie Universit\"at Berlin, 14195 Berlin, Germany}}

\author{Katharina J. Franke}
\affiliation{\mbox{Fachbereich Physik, Freie Universit\"at Berlin, 14195 Berlin, Germany}}

\author{Felix von Oppen}
\affiliation{\mbox{Dahlem Center for Complex Quantum Systems and Fachbereich Physik, Freie Universit\"at Berlin, 14195 Berlin, Germany}}

\date{Oct.\ 29, 2022}

\begin{abstract}
Current-biased Josephson junctions
exhibit hysteretic transitions between dissipative and superconducting states as characterized by switching and retrapping currents. Here, we develop a theory for diode-like effects in the switching and retrapping currents of weakly-damped Josephson junctions. We find that while the diode-like behavior of switching currents is rooted in asymmetric current-phase relations, nonreciprocal retrapping currents originate in asymmetric quasiparticle currents. These different origins also imply distinctly different symmetry requirements. We illustrate our results by a microscopic model for junctions involving a single magnetic atom.
Our theory provides significant guidance in identifying the microscopic origin of nonreciprocities in Josephson junctions. 
\end{abstract}

%\pacs{}

\maketitle 

{\em Introduction.---}The nonreciprocal behavior of diodes constitutes a central element of electronics \cite{Scaff1947,Shockley1949}. Nonreciprocity is also central to microwave and radio-frequency technology \cite{Pozar2011}. It is clearly a question of both applied and fundamental concern, whether nonreciprocal behavior can be realized in superconductors.
Recent experiments on superconductors \cite{Wakatsuki2017,Ando2020,Narita2022,Lyu2021,Shin2021,Lin2022,Hou2022,Narita2022,Sundaresh2022} as well as related theory \cite{Tokura2018,Daido2022,He2022,Ilic2022,Yuan2022} indicate that the critical current can indeed depend on the current direction.

\begin{figure}[t]
    \centering
    \includegraphics{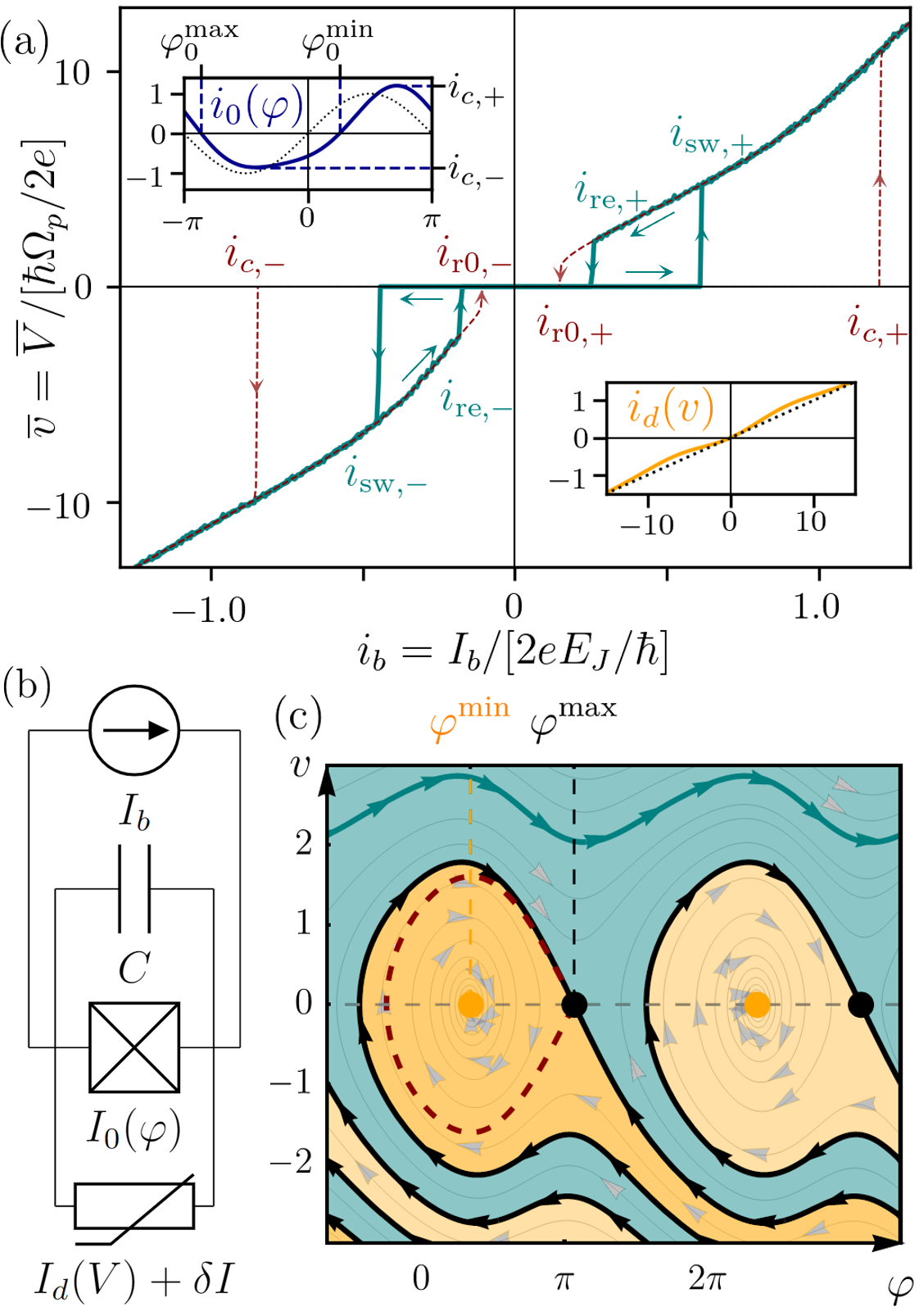}
    \caption{
    RCSJ model for weakly damped Josephson junctions. (a) Hysteretic dependence of time-averaged voltage on bias current  ($T=0$: red, dashed;  $T\neq 0$: green) and characteristic currents. Trace generated for asymmetric current-phase relation $I_0(\varphi)$ (left inset) and dissipative current  $I_d(V)$ (right inset). Dotted traces in insets show corresponding curves used in the conventional RCSJ model. (b) Equivalent circuit of the RCSJ model. (c) Phase-space diagram of the deterministic junction dynamics with coexisting trapped (orange) and running (green) solutions.  Parameters: \cite{supp}.
    }
\label{fig:RCSJ}
\end{figure}

These experiments have been extended to Josephson junctions, which are particularly promising for device applications for instance in the context of superconducting qubits. A variety of current-biased junctions have been found to exhibit nonreciprocal behavior \cite{Wu2022,Baumgartner2021,Diez2021,Bauriedl2022,Pal2022,Jeon2022,Turini2022,Gupta2022,Chiles2022}.
Many junctions are in the weak-damping regime, where the voltage response is hysteretic (Fig.\ \ref{fig:RCSJ}a) and the nonreciprocal behavior can occur in multiple characteristic currents. When increasing the bias current, the junction switches into the resistive state at the switching current $I_\textrm{sw}$. Conversely, when reducing the current bias, the junction will retrap into the supercurrent state at a smaller retrapping current $I_\textrm{re}$. The switching and retrapping currents are in general different from one another and from the critical current $I_c$ of the junction, the maximal supercurrent that the junction can in principle support \cite{Tinkham}. While the dominant nonreciprocity is typically in the switching current \cite{Wu2022,Pal2022,Turini2022,Chiles2022}, it is in the  retrapping current in a recent experiment \cite{Trahms2022}.

Theoretical work \cite{Hu2007,Yokoyama2013,Dolcini2015,Misaki2021,Zhang2021,Kopasov2021,Davydova2022,Souto2022,Tanaka2022}
has largely focused on exploring scenarios in which the current-phase relation and hence the critical current are asymmetric. Here, we present a general discussion of nonreciprocities in the various characteristic currents of current-driven Josephson junctions. Focusing on the low-damping limit with well-developed hysteresis, we show that nonreciprocities in the switching and retrapping currents have different microscopic origins and require different sets of broken symmetries. While dominant nonreciprocity in the switching current results from an asymmetric current-phase relation, dominant nonreciprocity in the retrapping current originates in asymmetric quasiparticle dissipation. We illustrate our results by a microscopic calculation for junctions including a single magnetic atom. Our results give important guidance for the design and interpretation of experiments on nonreciprocal Josephson junctions. 

{\em Model.---}The dynamics of Josephson junctions is conventionally described within the model of a resistively and capacitively shunted Josephson junction (RCSJ) \cite{Tinkham}. This model assumes that the junction carries capacitive ($I_C$), dissipative ($I_d$), and noise ($\delta I$) currents in parallel to the supercurrent ($I_0$). Current conservation implies that these currents sum to the bias current $I_b$ (Fig.\ \ref{fig:RCSJ}b), 
\begin{equation}
    I_C + I_0 + I_d + \delta I = I_b.
    \label{eq:RCSJ}
\end{equation}
The conventional RCSJ model  assumes a sinusoidal current-phase relation $I_0 = I_c \sin \varphi$, Ohmic dissipation $I_d=V/R$ by a shunt resistance $R$, and a capacitive current $I_C=C\dot V$. Johnson-Nyquist noise associated with the shunt resistor introduces a fluctuating current with correlator $\langle \delta I(t_1)\delta I(t_2)\rangle = \frac{2T}{R}\delta(t_1-t_2)$ at temperature $T$. The Josephson relation $V=\hbar \dot \varphi/2e$ turns Eq.\ (\ref{eq:RCSJ}) into a Langevin equation for the stochastic dynamics of the superconducting phase difference $\varphi$ across the junction. 

In its conventional form, the RCSJ model predicts reciprocal characteristic currents. Nonreciprocal behavior can in general be introduced by modified capacitive, dissipative, or  supercurrent terms. Misaki and Nagaosa \cite{Misaki2021} showed that  nonreciprocal behavior can originate in nonlinear contributions to the quantum capacitance. This mechanism requires different carrier densities on the two sides of the junction and 
applies to junctions joining two different superconducting materials. 

The recent experiments \cite{Wu2022,Pal2022, Turini2022,Chiles2022,Trahms2022} 
were performed on junctions made of a single superconductor. For this reason, we concentrate on nonreciprocities originating in the supercurrent $I_0$ and the dissipative current $I_d$. To this end, we allow for general current-phase relations
$I_0(\varphi)$ and dissipative currents $I_d = I_d(V)$. Using Eq.\ (\ref{eq:RCSJ}) and the Josephson relation, the phase dynamics is then described by \cite{Stewart1969,Kautz1990}
\begin{equation}
    (\hbar C/2e)\ddot \varphi + I_d (\hbar\dot \varphi/2e)
    + I_0(\varphi)  +\delta I = I_b.
    \label{eq:Langevin_dim}
\end{equation}
The correlator $\langle \delta I(t_1)\delta I(t_2)\rangle = K(V=\hbar\dot \varphi/2e) \delta(t_1-t_2)$ of the current fluctuations is related to the dissipative current by the fluctuation-dissipation theorem. In the limit of low temperatures, this implies $K(V)=2T\frac{I_d(V)}{V}$ (for a detailed discussion, see the Supplementary Material \cite{supp}). Equation (\ref{eq:Langevin_dim}) describes the dissipative motion of a phase particle in a tilted washboard potential $U(\varphi)=U_0(\varphi) -(\hbar/2e) I_b\varphi$ with $I_0(\varphi)=(2e/\hbar)({dU_0}/{d\varphi})$. For definiteness, we restrict our attention to (periodic) potentials $U_0(\varphi)$ with a single minimum ($\varphi^{\textrm{min}}_{0}$) and maximum ($\varphi^{\textrm{max}}_{0}$) per period.

\begin{figure}[t]
    \centering
    \includegraphics{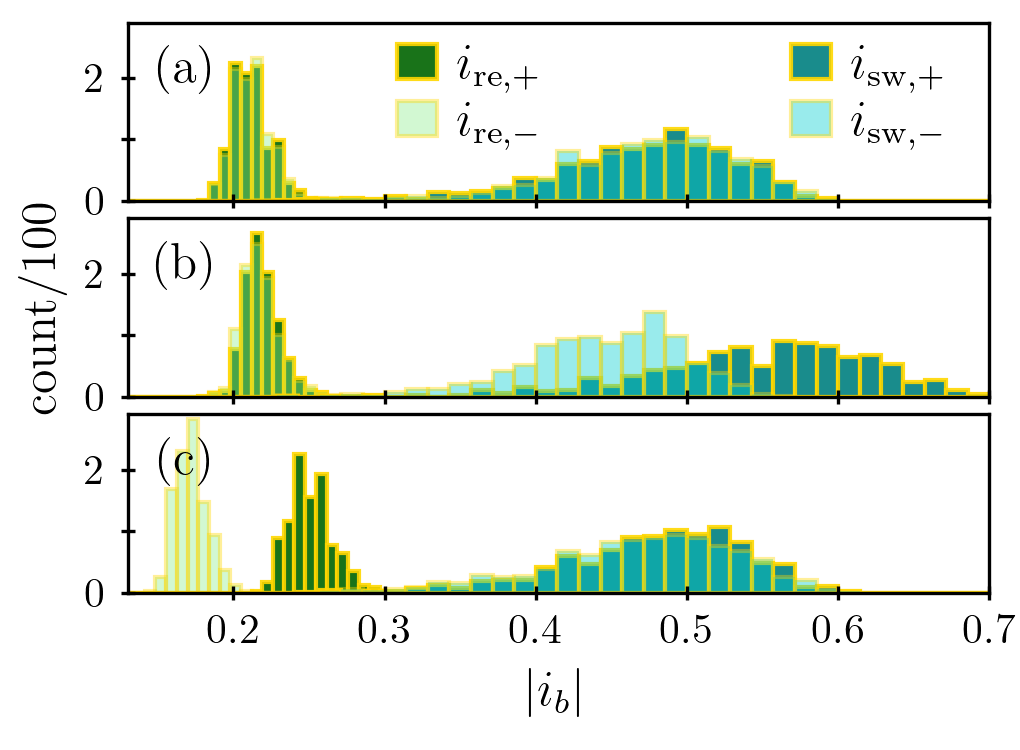}
    \caption{
    Histograms of retrapping ($i_{\mathbf{re,\pm}}$) and switching ($i_{\mathbf{sw,\pm}}$) currents for bias currents $i_b$ of both signs ($\pm$). (a) Conventional RCSJ model. (b) Asymmetric current-phase relation $I_0(\varphi)$ and symmetric dissipative current $I_d(\varphi)$. (c) Symmetric $I_0(\varphi)$ and asymmetric $I_d(\varphi)$. Parameters: \cite{supp} }
\label{fig:histograms}
\end{figure}

{\em Nonreciprocity and symmetries.---}The  nonreciprocity in the switching and retrapping currents have distinctly different origins. This can be seen by directly simulating the Langevin dynamics for a weakly damped junction. Resulting histograms for the switching and retrapping currents are shown in Fig.\ \ref{fig:histograms}. The histograms are independent of the direction of the bias current $I_b$ for the conventional RCSJ model (Fig.\ \ref{fig:histograms}a). Only the switching current is nonreciprocal, when the junction has an asymmetric current-phase relation, $I_0(\varphi) \neq - I_0(-\varphi)$, but  symmetric dissipative current, $I_d(V)=-I_d(-V)$. In contrast, only the retrapping current is nonreciprocal for asymmetric $I_d(V)$, but symmetric $I_0(\varphi)$. 

This difference between switching and retrapping currents reflects that switching and retrapping are due to different underlying physics. Switching is caused by  escape from a minimum of the tilted washboard potential $U(\varphi)$, thus requiring asymmetry in the $U_0(\varphi)$ and hence $I_0(\varphi)$. In contrast, retrapping back into a minimum of $U(\varphi)$ is induced by frictional energy loss, which depends directly on the dissipative current $I_d(V)$ for the particular bias direction. This also implies that nonreciprocities in the switching and retrapping currents have different symmetry requirements. Asymmetries in the current-phase relation require breaking of both time-reversal and inversion symmetry \cite{Wakatsuki2017,Davydova2022}. In contrast, asymmetries of the dissipative current, $I_d(V) \neq - I_d(-V)$, require breaking of particle-hole symmetry (in the normal-metal sense) and inversion symmetry, while time reversal need not be broken (as for conventional diodes). The dissipative current has contributions from the quasiparticle current of the junction as well as the electromagnetic environment. While the latter is typically symmetric, the quasiparticle current is generically nonlinear and asymmetric in the absence of inversion and particle-hole symmetry.

%This also implies that nonreciprocities in the switching and retrapping currents have different symmetry requirements. Asymmetries in the current-phase relation require breaking of both time-reversal and inversion symmetry. Microscopic time-reversal symmetry implies $F(\varphi,B) = F(-\varphi,-B)$ for the free energy as a function of $\varphi$ and applied magnetic field $B$. Consequently, the supercurrent $I_0=({2e}/{\hbar})({\partial F}/{\partial\varphi})$ obeys $I_0(\varphi,B) =  -I_0(-\varphi,-B)$. In the presence of time-reversal symmetry, i.e., for $B=0$, this yields a symmetric current-phase relation $I_0(\varphi) = -I_0(-\varphi)$. Similarly, inversion symmetry of the junction implies $F(\varphi,B) =   F(-\varphi,B)$ and thus $I_0(\varphi,B) =   -I_0(-\varphi,B)$ even without time-reversal symmetry. 

%Asymmetries of the dissipative current, $I_d(V) \neq - I_d(-V)$, require breaking of particle-hole symmetry (in the normal-metal sense) and inversion symmetry, while time reversal need not be broken (as for conventional diodes). The dissipative current has contributions from the quasiparticle current of the junction as well as the electromagnetic environment. While the latter is typically symmetric, the quasiparticle current is generically nonlinear and asymmetric in the absence of inversion and particle-hole symmetry.

{\em Fokker-Planck description.---}To develop an analytical theory, we follow standard considerations to convert Eq.\  (\ref{eq:Langevin_dim}) into the Fokker-Planck equation \cite{Ambegaokar1969,Haenggi1990}
\begin{equation}
    \frac{\partial p}{\partial\tau} =
    \left(-v \frac{\partial}{\partial \varphi}
    +\frac{\partial}{\partial v}
    \left[u'(\varphi) + i_d(v) 
    +\frac{1}{2}
    \frac{\partial}{\partial v}k(v)\right]
    \right)
    p
\end{equation}
for the time evolution of the probability density $p(\varphi,v;\tau)$ as a function of the phase $\varphi$ and its velocity $v=\varphi'$. Here, we have defined a dimensionless time variable $\tau = \Omega_p t$ in terms of the plasma frequency $\Omega_p=[4e^2 E_J/\hbar^2 C]^{1/2}$, where 
$E_J=d^2 U_0(\varphi^{\textrm{min}}_{0})/d\varphi^2$ is the Josephson energy.
(This implies $U_0(\varphi)=-E_J\cos\varphi$ for a sinusoidal current-phase relation.) Also defining dimensionless currents $i = (\hbar/2e)I/E_J$ and  potentials $u = U/E_J$, the Langevin equation (\ref{eq:Langevin_dim}) becomes $\varphi'' + i_d( \varphi') + i_0(\varphi) +\delta i = i_b$,
where primes denote derivatives with respect to dimensionless time $\tau$. The noise correlator $\langle \delta i(\tau_1)\delta i(\tau_2)\rangle = k ( \varphi')\delta(\tau_1-\tau_2)$ involves $k(v) = 2\theta i_d(v)/v$ in terms of the reduced temperature $\theta = T /E_J$. 

At zero temperature, the Johnson-Nyquist noise vanishes, $\delta i = 0$. Then, the dynamics of the phase variable becomes deterministic, with two types of solutions. For small bias currents, the phase is locked to a minimum $\varphi^{\textrm{min}}$ of the washboard potential $u(\varphi)$ and the junction supports supercurrent flow, $i_b = i_0(\varphi^{\textrm{min}})$. For large bias currents, there is a running solution corresponding to a resistive state of the junction. 
In this state, the phase variable moves in a fixed direction at all times and the energy gain due to the current bias is compensated by the friction induced by the dissipative current.  

For weak damping, the two types of solutions coexist at intermediate bias currents, see the phase-space diagram in Fig.\ \ref{fig:RCSJ}(c). Then, the junction transitions between the two types of solutions due to Johnson-Nyquist noise and exhibits hysteresis. The nonreciprocal behavior of Josephson junctions is controlled by the transition rates between the trapped and running states, which we now derive for general current-phase relations and dissipative currents.

{\em Switching rate.---}We first consider the switching rate out of the trapped into the running state. For weak damping, the energy - and consequently the action -- of the undamped motion are  slowly-varying variables. Then, the Fokker-Planck equation can be reduced to a drift-diffusion equation for the distribution function $p(J)$ of the action $J=\oint d\varphi\, v$ (here, $\oint$ denotes an integral over one period of the trapped motion) \cite{Kramers1940,Lee1971}. Using the general drift-diffusion equation
$  \partial_\tau p = \partial_J [ - v_D p + D \partial_J p]$ and deducing the drift velocity $v_D = -\oint d\varphi\, i_d(v)$ as well as the diffusion coefficient $D = \frac{2\pi\theta}{\omega(J)} \oint d\varphi\, i_d(v)$ from the Langevin equation, we obtain (see \cite{supp} for details)
\begin{equation}
    \frac{\partial p}{\partial \tau} =   \frac{\partial}{\partial J} \left\{ \varepsilon_d(J) \left[ 1 +
    \frac{2\pi\theta}{\omega(J)}  
    \frac{\partial}{\partial J}
    \right]\right\} p.
    \label{eq:dd_sw}
\end{equation}
Here, we introduced the (dimensionless) energy $\varepsilon_d = \oint d\varphi\, i_d(v)$, which is dissipated per period. The current-phase relation and the bias current also enter via the angular frequency $\omega(J)=2\pi\,  dh(J)/dJ$ of the trapped motion, where $h=\frac{1}{2}v^2+u(\varphi)$ is the Hamiltonian of the undamped junction.

\begin{figure*}[ht]
    \centering
    \includegraphics{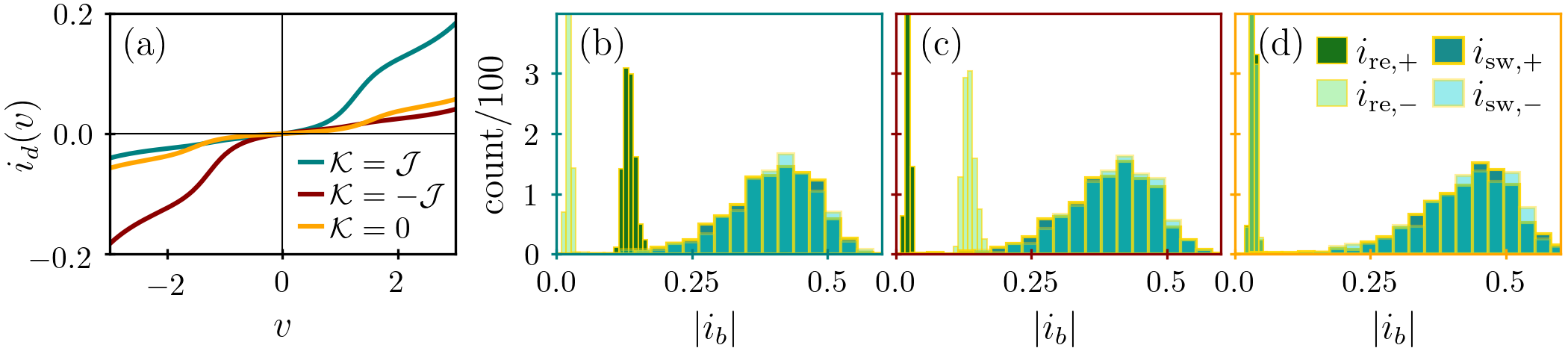}
    \caption{ Microscopic model of Josephson junction with magnetic impurity. (a) $I$-$V$ characteristics of  quasiparticle current due to YSR state associated with magnetic impurity for different potential scatterings $\mathcal{K}$. (b-d) Histograms of switching and retrapping currents corresponding to $I$-$V$ characteristics in (a) (note color-coded box), emphasizing the importance of particle-hole symmetry. Parameters: \cite{supp}.}
    \label{fig:YSR}
\end{figure*}

Deriving the transition rate from the trapped into the running state is now an escape problem out of a metastable well \cite{Haenggi1990}. In the low-temperature limit, we find the (dimensionless) activation rate (see \cite{supp})
\begin{equation}
    \gamma_\mathrm{sw}
    = \frac{\varepsilon_\textrm{d}(J_B) \omega_0}{2\pi \theta}
    \exp\left\{-\frac{\varepsilon_B}{\theta}\right\}.
    \label{gswi}
\end{equation} 
The transition rate depends exponentially on the activation barrier $\varepsilon_B=h(J_B)-h(J=0)$ out of the metastable well. (Here, $J_B$ is the action of the undamped separatrix motion beginning and ending at  the unstable maximum $\varphi^\mathrm{max}$, see red dashed contour in Fig.~\ref{fig:RCSJ}(c), and $J=0$ corresponds to the phase particle at rest in the stable minimum $\varphi^\mathrm{min}$.) 
Dissipation only affects the preexponential attempt frequency through $\varepsilon_d(J_B)$, the limit of the dissipated energy per period as the separatrix between running and trapped motion is approached. $\omega_0$ is the oscillation frequency in the minimum of the tilted washboard potential $u(\varphi)$. 

{\em Retrapping rate.---}We now consider the  retrapping rate from the running into the trapped state. For weak damping, the retrapping current is parametrically smaller than the critical current and proportional to the strength of dissipation. We can thus restrict attention to  small bias currents implying a weakly-tilted washboard potential. Under these conditions,  we focus on  the action $J=\int_0^{2\pi\, \mathrm{sgn}(v)} d\varphi\, v$ evaluated for the Hamiltonian $h_0=\frac{1}{2}v^2+u_0(\varphi)$ of the unbiased junction. Note that the action $J$ is now defined as an integral over all $\varphi$,
as appropriate for the running state. The action is again slowly varying with time and satisfies a drift-diffusion equation \cite{Cristiano1986}. The drift  involves a contribution from the bias current $i_b$ in addition to the dissipative term, $    v_d = -\int_0^{2\pi\, \mathrm{sgn}(v)} d\varphi\, i_d(v) + 2\pi |i_b|$. The diffusion constant becomes $  D = \frac{2\pi\theta}{\omega(J)} \int_0^{2\pi\, \mathrm{sgn}(v)} d\varphi\, i_d(v)$. This gives \cite{supp}
\begin{equation}
    \frac{\partial p}{\partial \tau} =   \frac{\partial}{\partial J} \left\{ \varepsilon_d(J)   -2\pi |i_b| + 
    \varepsilon_d(J)\frac{2\pi\theta}{\omega(J)}  
    \frac{\partial}{\partial J} \right\} p
    \label{eq:dd_re}
\end{equation}
for $p=p(J)$. Here, $\varepsilon_d(J) = \int_0^{2\pi\, \mathrm{sgn}(v)} d\varphi\, i_d(v)$ is the dissipated energy per period in the running state. 

Following analogous steps as for the switching rate, we obtain the (dimensionless) retrapping rate \cite{supp} 
\begin{eqnarray}
    \gamma_\mathrm{re}
    &=& \sqrt{\frac{i_d^\prime(\overline{v})}{i_d(\overline{v})/\overline{v}}\frac{[|i_b|-i_{\mathrm{r}0,\pm}]^2}{2\pi\theta}}
    \nonumber\\
    &&\times
    \exp\left\{-
    \frac{1}{2\theta}\frac{1}{ i_d^\prime(\overline{v}) \frac{i_d(\overline{v})}{\overline{v}}} [|i_b| - i_{\mathrm{r}0,\pm}]^2\right\}.
\end{eqnarray}
Here, we define the retrapping currents $i_{r0,\mathrm{sgn}v} = \frac{1}{2\pi}\int_0^{2\pi\, \mathrm{sgn}(v)}d\varphi\, i_d(v(J_B))$ in the absence of fluctuations, where $J_B$ is the action of the separatrix beginning and ending at neighboring unstable maxima $\varphi^\mathrm{max}_0$. The average phase velocity $\overline v$ is the solution of  $i_b=i_d(\overline{v})$ and the upper (lower) sign applies for $i_b>0$ ($i_b<0$). The expression for the retrapping rate is valid at low temperatures and for bias currents sufficiently far from  $i_{\mathrm{r}0,\pm}$. For Ohmic friction and sinusoidal current-phase relation, our result reduces to the classic expression of Ben-Jacob {\em et al.} \cite{Benjacob1982}.

{\em Nonreciprocity of switching and retrapping currents.---}With these preparations, we are in a position to discuss  nonreciprocity in weakly-damped Josephson junctions in rather general terms. First consider the nonreciprocity properties of the switching rate. If $u_0(\varphi)$ is symmetric about $\varphi=0$ [and thus $i_0(\varphi)=-i_0(-\varphi)$], Eq.\ (\ref{eq:dd_sw}) governing the switching rate is explicitly symmetric under sign changes of the bias current $i_b$. This follows since the dissipation parameter $\varepsilon_d$ and the frequency $\omega(J)$ are expressed as integrals over a full period of the trapped motion, in which the phase velocity $v$ changes sign \cite{supp}. 
This conclusion remains true even if the dissipative current $i_d(v)$ is asymmetric in $v$. We thus find that as for the critical current, nonreci\-procal switching rates require breaking of time reversal symmetry, such that $u_0(\varphi)$ is no longer symmetric under $\varphi\to -\varphi$ and the barrier $\varepsilon_B$ becomes dependent on the sign of the bias current $i_b$.

Equation (\ref{eq:dd_re}) governing the
retrapping rate is explicitly symmetric under sign changes of $i_b$ as long as the dissipative current $i_d(v)$ is symmetric. This is true for any potential $u_0(\varphi)$. A nonreciprocal retrapping rate originates in asymmetry of $i_d(v)$, which leads to $i_{r0,+}\neq i_{r0,-}$. Thus, a nonreciprocal retrapping rate requires breaking of  particle-hole symmetry (in the normal-metal sense), so that the contribution of the  quasiparticle current to $i_d(v)$ can  be asymmetric. In contrast, breaking of time-reversal  symmetry [asymmetric $u_0(\varphi)$] is not required. 

The switching and retrapping rates allow one to derive expressions for the average switching and retrapping currents. We assume a bias-current ramp $i_b(\tau)=a \tau$ with rate $a$. Then the switching current $i_\mathrm{sw}$ can be defined through $P_t(i_\mathrm{sw})=\frac{1}{2}$, where $P_t$ is the probability to be in the trapped state. Using the switching rate (\ref{gswi}), one finds that the shift in the switching current $\Delta i_{\mathrm{sw},\pm} = i_{\mathrm{sw},\pm}- i_{c,\pm}$ relative to the critical current is equal to \cite{supp}
\begin{equation}
    \frac{\Delta i_{\mathrm{sw},\pm}}{i_{c,\pm}} \approx
    \Bqty{
    \frac{\theta}{ \varepsilon_{B0} } \ln \bqty{
   \frac{ \varepsilon_d(J_{B0})}{2\pi a \ln 2 \abs{\varphi_0^{\textrm{max}} - \varphi_0^{\textrm{min}} }_{\pm}}
   }
   }^{\frac{1}{\mu_\pm}},
   \label{eq:iswfl}
\end{equation}
where $J_{B0}$ is the action of the separatrix in the absence of damping and bias current and $\mu_\pm= \abs{\varphi_0^{\textrm{max}} - \varphi_0^{\textrm{min}} }_{\pm} i_{c,\pm}/\varepsilon_{B0}$,. Similarly, we find that fluctuations shift the retrapping rate away from $i_{\mathrm{r}0,\pm}$ by \cite{supp}  \begin{equation}
    \Delta i_{\mathrm{re},\pm} \approx  
    \Bqty{\theta\   i_d^\prime(\overline{v}) \frac{i_d(\overline{v})}{\overline{v}} \ln \bqty{
    \frac{\theta  \pqty{i_d^\prime(\overline{v})}^3 }{ 2\pi (a \ln 2)^2}  \frac{i_d(\overline{v})}{\overline{v}}
    } }^{\frac{1}{2}},
    \label{eq:irefl}
\end{equation}
where the right hand side is evaluated for $i_b = \pm i_{\mathrm{r}0,\pm} $. These expressions make the nonreciprocities of the average switching and retrapping currents explicit. While Eqs.\ (\ref{eq:iswfl}) and (\ref{eq:irefl})
assume sufficiently high barriers as well as smooth drift and diffusion of the action \cite{supp}, our qualitative results are valid more widely as indicated by the numerical results in Fig.\ \ref{fig:YSR}. We also note that while our analytical results focus on the regime of thermally activated switching and retrapping, quantum tunneling may become relevant at sufficiently low temperatures. This will affect explicit temperature dependences, but leaves our qualitative results unaffected.

{\em Yu-Shiba-Rusinov junctions.---}We illustrate the importance of particle-hole symmetry for the retrapping current by   microscopic results for a Josephson junction hosting a magnetic adatom coupled to one of the electrodes (for a recent experiment, see \cite{Trahms2022}). The spin of the adatom couples to the electrode electrons via both exchange scattering $\mathcal{J}$ and potential scattering $\mathcal{K}$, with the latter being nonzero only when particle-hole symmetry is broken \cite{Schrieffer1966}.
These couplings induce Yu-Shiba-Rusinov (YSR) resonances within the superconducting gap, which are symmetric in energy, but in general asymmetric in intensity for nonzero potential scattering. This provides a microscopic model for an asymmetric quasiparticle current (Fig.\ \ref{fig:YSR}a) accompanied by a symmetric current-phase relation. Consistent with our general theoretical analysis, a simulation of the junction dynamics based on a standard model for YSR states \cite{Yu1965,Shiba1968,Rusinov1969,Leshouches2017} (see \cite{supp} for details) exhibits asymmetric retrapping currents for nonzero  potential scattering $\mathcal{K}$, with the direction of the asymmetry dependent on the sign of $\mathcal{K}$ (Fig.\ \ref{fig:YSR}b and c). Symmetric retrapping currents are observed for $\mathcal{K}=0$, when particle-hole symmetry is preserved (Fig.\ \ref{fig:YSR}d).

{\em Conclusions.---}We have developed a general theory of nonreciprocity in current-biased Josephson junctions, focusing on the hysteretic behavior for weak dissipation (high quality factor). We have shown that a nonreciprocal switching current originates from nonreciprocity of the supercurrent. In contrast, a nonreciprocal retrapping current originates from  quasiparticle dissipation which is asymmetric under a sign change of $i_b$. Moreover, these different sources of nonreciprocity have different symmetry requirements. While nonreciprocal switching currents require breaking of time reversal symmetry, nonreciprocity of the retrapping current requires breaking of particle-hole symmetry, but not of time reversal symmetry. Recent experiments on weakly damped Josephson junctions revealed  dominant  nonreciprocities in both, the switching and the retrapping current. Our theory implies that these nonreciprocities have fundamentally different microscopic origins. 

\begin{acknowledgments}
We gratefully acknowledge funding by Deutsche Forschungsgemeinschaft through CRC 183 (project C03) as well as CRC 910 (project A11). 
\end{acknowledgments}

%\bibliographystyle{apsrev4-2}
%\bibliography{library}


\newpage

\appendix

\begin{widetext}

\clearpage

\setcounter{figure}{0}
\setcounter{section}{0}
\setcounter{table}{0}
\setcounter{section}{0}
\renewcommand{\thefigure}{S\arabic{figure}}
\renewcommand{\thetable}{S\arabic{table}}
\renewcommand{\thesection}{S\arabic{section}}
\renewcommand{\theequation}{\thesection.\arabic{equation}}

\onecolumngrid

\newcommand{\vsigma}{\mbox{\boldmath $\sigma$}}

\section*{\Large{Supplementary Information}}

\section{Numerical simulation and parameter values}

All our Langevin simulations employ the Euler-Maruyama algorithm with time step $d\tau = 0.01$. In order to obtain hysteretic behaviour, we sweep the bias current from $i_b = - 0.7$ to $i_b = 0.7$ [$i_b = - 1.5$ to $i_b = 1.5$ in Fig.~\ref{fig:RCSJ}a and $i_b = - 0.9$ to $i_b = 0.9$ in Fig.~\ref{fig:histograms}b] and back at a rate $r = \abs{di/d\tau} = 10^{-5}$. The time-averaged voltage is calculated for $500$ points in each direction. Switching and retrapping currents are extracted as maxima of the discrete derivative of the time-averaged voltage data. For the histograms we run $1000$ sweeps.

In Figs.\ \ref{fig:RCSJ} and \ref{fig:histograms} we use the phenomenological current-phase relation 
\begin{equation}
    i_0(\varphi) = c_1 \bqty{ \sin(\varphi - \tilde{\varphi}) - c_2 \sin(2\varphi) },
\end{equation} 
where $c_1$ is fixed such that $(\partial^2 u_0/\partial \varphi^2)\vert_{\varphi^\mathrm{min}_0} = (\partial i_0/\partial \varphi)\vert_{\varphi^\mathrm{min}_0} = 1$, see the top inset in Fig.\ \ref{fig:RCSJ}a. This type of current-phase relation arises, e.g., in Ref.\ \cite{Tanaka2022}. For the dissipative current we use the current-voltage characteristic
\begin{equation}
    i_d(v) = \frac{v}{Q} \bqty{1 + c_3 \frac{v}{\delta v} \exp{-\frac{1}{2}\pqty{\frac{v^2}{\delta v^2} - 1}}},
\end{equation}
see the bottom inset in Fig.\ \ref{fig:RCSJ}a. The exponential factor in the asymmetric term serves to give Ohmic behavior at large voltages. As parameters we use $Q=10$ and $\theta  = 0.1$ throughout Figs.\ \ref{fig:RCSJ} and \ref{fig:histograms}. For the asymmetric current-phase relation [Figs.\ \ref{fig:RCSJ}a and \ref{fig:histograms}b] we use $
\tilde{\phi} = 0.6$ and $c_2/c_1 = 0.2$. For the asymmetric current-voltage characteristic [Figs. \ref{fig:RCSJ}a and \ref{fig:histograms}c] we use $c_3 = 0.3, \delta v = 5$. For the symmetric current-phase relation [Figs.\ \ref{fig:histograms}a and \ref{fig:histograms}c] we use $c_2 = 0 = \tilde{\varphi}$ s.t. $i_0 \sin\varphi$. For the symmetric current-voltage characteristic [Figs.\ \ref{fig:histograms}a and \ref{fig:histograms}b] we use $c_3 = 0 = \delta v$, s.t. $i_d(v) = v/Q$. Finally, in Fig.\ \ref{fig:RCSJ}(c) the bias current is $i_b = 0.3$. 

In Fig. \ref{fig:YSR} we use the following parameters: tip and substrate gap $\Delta = 2\hbar\Omega_p$, temperature $T = 0.01\Delta$, dimensionless exchange scattering $\alpha = \pi\nu_0 \mathcal{J} = 1.5$, dimensionless tip-substrate coupling $\gamma = \pi\nu_0\mathcal{T} = 0.2$, broadening $\eta = 0.1\Delta$ and dimensionless potential scattering $\beta = \pi\nu_0\mathcal{K} \in \{\pm \alpha, 0\}$ (for details see App. \ref{sec:ysr} below).   

\section{Deterministic junction dynamics}
\label{sec:determ}

We review the deterministic junction dynamics in the absence of the Langevin term (Fig.\ 1c of the main text). We systematically work in dimensionless units. With the exceptions of time $t$ (dimensional) and action $J$ (dimensionless), lower (upper)-case quantities are in dimensionless (dimensional) units. The phase-space trajectories obey the equations of motion
\begin{align}\label{eq:phase_space_traj}
    \frac{d\varphi}{d\tau}= v \qquad ,\qquad 
    \frac{dv}{d\tau} = -u'(\varphi) - i_d(v). 
\end{align}
Combining the two equations of motion, we obtain
\begin{equation}
     \frac{d }{d\varphi}\left[\frac{1}{2}v^2 + u(\varphi)\right] = - i_d(v)
     \label{eq:phase_space}
\end{equation}
for the  phase-space trajectories $v=v(\varphi)$ as illustrated in Fig.\ 1c of the main text.
In the absence of fluctuations, trapped solutions correspond to the stable fixed point $\varphi = \varphi^{\textrm{min}}$ and $v = 0$ in phase space  (orange point in Fig.\ 1c). The running solution (thick green trace in Fig.\ 1c) is the only periodic solution $v=v(\varphi)$ of Eq.\ (\ref{eq:phase_space}), for which the dissipated energy per period is balanced by the energy provided by the bias current, 
\begin{equation}
    2\pi i_b = \int_0^{2\pi} d\varphi\,  i_d(v).
\end{equation}

Neglecting fluctuations, the switching currents $i_{s0,\pm}$ are equal to the critical currents $i_{c,\pm}$, at which the minima of the tilted washboard potential cease to exist. In the weak-damping limit, the retrapping current can be obtained as follows. We  consider 
the phase-space trajectories $v(\varphi)$ of the running solutions in the absence of bias current and damping, which obey
\begin{align}
     h_0  =\frac{1}{2}v^2 + u_0(\varphi),
\end{align}
with the reduced energy $h_0$ of the phase-space trajectory. Below (above) the energy $\varepsilon_B =  u_0(\varphi_0^\textrm{max})$, all trajectories are trapped (running). The separatrix motion between these two classes of trajectories begins and ends at neighboring unstable fixed points $(\varphi_0^\mathrm{max},v=0))$. As a consequence of the time reversal symmetry of the undamped motion, there are two separatrix solutions $v^\pm(\varphi)$ with opposite signs of the velocity. Now, the retrapping current follows by equating the energy dissipation along the separatrix to the energy gain due to the bias current, 
\begin{equation}
    i_{\mathrm{r}0,\pm} = \frac{1}{ 2\pi}\int_0^{\pm 2\pi} d\varphi\, i_d(v^\pm(\varphi)).
    \label{eq:retrap_det}
\end{equation}
Note that the retrapping currrent is proportional to the strength of the dissipation and thus parametrically small compared to the switching current. Importantly, the retrapping current is asymmetric in the bias direction for asymmetric dissipative currents, $i_d(v)\neq -i_d(-v)$.

\section{Stationary solution of the Fokker-Planck equation and detailed balance}

Detailed balance implies a relation between $k(v)$ and $i_d(v)$ in the Fokker-Planck equation 
\begin{equation}
    \frac{\partial p}{\partial\tau} =
    \left(-v \frac{\partial}{\partial \varphi}
    +\frac{\partial}{\partial v}
    \left[u'(\varphi) + i_d(v) 
    +\frac{1}{2}
    \frac{\partial}{\partial v}k(v)\right]
    \right)
    p
\end{equation}
for $p=p(\varphi,v;\tau)$ given in the main text. There, we quoted this relation in the limit of low temperatures. Here, we give a more general derivation. 

The Fokker-Planck equation can be viewed as a continuity equation in phase space
$\partial_\tau p= -\nabla\cdot \mathbf{j}$
with the current density
\begin{equation}
    j_\varphi = vp
     \quad , \quad j_v = -\left[u'(\varphi)+i_d(v)+\frac{1}{2}
    \frac{\partial}{\partial v}k(v)\right]p.
\end{equation}
We separate the current density into a reversable and an irreversable contribution,
$\mathbf{j} = \mathbf{j}^\mathrm{rev} + \mathbf{j}^\mathrm{irr}$, with
\begin{equation}
   j^\mathrm{rev}_\varphi = vp 
   \quad , \quad j^\mathrm{rev}_v = 
   -u'(\varphi)p
\end{equation}
and
\begin{equation}
   j^\mathrm{irr}_\varphi = 0 
   \quad , \quad j^\mathrm{irr}_v = 
   -\left[i_d(v)+\frac{1}{2}
    \frac{\partial}{\partial v}k(v)\right]p.
\end{equation}
The stationary solution $p(\varphi,v)$ satisfies $\nabla\cdot \mathbf{j}=\mathbf{0}$. In contrast to  the one-dimensional Fokker-Planck equations, 
we cannot conclude that $\mathbf{j}=\mathbf{0}$ in equilibrium. 
However, detailed balance requires that the irreversible contribution to the current  vanishes by itself, $\mathbf{j}^\mathbf{irr}=\mathbf{0}$. Thus, detailed balance demands
\begin{equation}
    \left[i_d(v)+\frac{1}{2}
    \frac{\partial}{\partial v}k(v)\right]p=0.
    \label{eq:detbal}
\end{equation}
This is easily integrated to give
\begin{equation}
    p(\varphi,v) = \frac{A(\varphi)}{k(v)}\exp\left\{
    -\int_0^v dv' \frac{2i_d(v')}{k(v')}
    \right\},
\end{equation}
where $A(\varphi)$ is an arbitrary integration constant. This solution can be written as 
\begin{equation}
    p(\varphi,v) =
    \exp{-[T(v)+U(\varphi)]}
\end{equation}
with 
\begin{equation}\label{eq:T(v)}
    T(v) = \int_0^v dv'\ \bqty{ \frac{2i_d(v') + k'(v')}{k(v')} }.
\end{equation}
Of course, this way of writing the solution is motivated by the Boltzmann distribution.  

Inserting the solution in this form into $\nabla\cdot \mathbf{j}^\mathrm{rev}=\mathbf{0}$ gives
\begin{equation}
    \frac{U'(\varphi)}{u'(\varphi)}
    = \frac{T'(v)}{v}.
    \label{eq:sep}
\end{equation}
This can only be satisfied if both sides of the equation are individually equal to a constant, say $\frac{1}{\theta}$. Integrating both sides, we find the expected expressions
\begin{equation}
    U(\varphi) = \frac{u(\varphi)}{\theta} + \mathrm{const},\ T(v) = \frac{v^2}{2\theta} + \mathrm{const},
\end{equation}
so that we can identify $\theta$ with temperature, reproducing the Boltzmann distribution.

Consistency with Eq.\ \eqref{eq:T(v)} implies 
\begin{equation}\label{eq:detailed_balance_eqn}
    T'(v) = \frac{v}{\theta} = \frac{2 i_d(v) + k'(v)}{k(v)}
\end{equation}
or 
\begin{equation}
    k'(v) - \frac{vk(v)}{\theta} + 2 i_d(v) =0.
\end{equation}
This equation is readily solved to give 
\begin{equation}
    k(v) =  2e^{v^2/2\theta}
    \int_v^\infty dv' e^{-v^{\prime 2}/2\theta} i_d(v'),
    \label{eq:kgen}
\end{equation}
which is the desired relation between $k(v)$ and $i_d(v)$. 
Here, we have specified an integration constant such that $k(v)$ reduces to the familiar result for the case of Ohmic friction.

In the limit of small $\theta$, this general result can be further evaluated as
\begin{equation}
        k(v) =  
    2\int_v^\infty dv' e^{-(v^{\prime 2}-v^2)/2\theta} i_d(v')
    \simeq 2\theta\frac{i_d(v)}{v},
\end{equation}
which is the expression given in the main text. 
Note that as advertised, this just involves the conductance in the usual way when specifying to the Ohmic limit. We note in passing that the choice of $k(v)$ in Ref.\ \cite{Kautz1990} is inconsistent with detailed balance.

\section{Escape from the trapped state}

\subsection{Fokker-Planck equation in weak-damping limit}

We now consider fluctuation-induced transitions out of the trapped state in the limit of weak friction. In the absence of friction and noise, we can describe the deterministic motion within the metastable minimum using action-angle variables $(J,w)$ \begin{equation}
    J = \oint d\varphi\, v = \textrm{const}
\end{equation}
and $w = \frac{\omega(J)}{2\pi} \tau + w_0$. The Hamiltonian $h=h(J)$ and the frequency of the trapped motion are related through $\frac{\omega(J)}{2\pi} = \frac{\partial h(J)}{\partial J}$. Note that here, we consider the trapped motion in the tilted washboard potential and the integrals are over one period of the periodic motion in the absence of friction, i.e. the Hamiltonian is given by
\begin{equation}
    h = \frac{1}{2}v^2 + u(\varphi).
\end{equation}

In the weak-friction limit, the angle is quickly varying, while the action variable changes only slowly. For this reason, one can average the motion over the fast phase variable and reduce the Fokker-Planck equation to an equation for a distribution function $p\simeq p(J)$, which depends only on the action. 
Friction along with the fluctuations will make the action drift and diffuse, so that the action satisfies a drift-diffusion equation
\begin{equation}
    \frac{\partial p}{\partial \tau} = \frac{\partial }{\partial J}\left[ - v_D p + D \frac{\partial p}{\partial J}
    \right].
\end{equation}
The diffusion coefficient $D=D(J)$ and the deterministic drift velocity $v_D=v_D(J)$ can themselves depend on the action. To determine the drift velocity, we consider the time derivative of the action,
\begin{equation}
    \dot J = \frac{\partial J}{\partial h} \dot h.
\end{equation}
This yields
\begin{equation}
    \dot J = \frac{\partial J}{\partial h} v \left[ 
    \dot v + u'(\varphi)
    \right]
    =-\frac{\partial J}{\partial h} v\, i_d(v) .
\end{equation}
Averaging over one period $\tau_0=2\pi/\omega(J)$, we obtain the drift velocity
\begin{equation}
    v_D = \langle \dot J\rangle_T = 
    \frac{1}{\tau_0} \int_0^{\tau_0} d\tau\, \dot J =
    -\frac{\omega(J)}{2\pi}\int_0^{\tau_0} d\tau\, \frac{\partial J}{\partial h} v i_d(v)  
    =- \oint d\varphi\, i_d(v)
\end{equation}
of the action.

The diffusion of the action arises from the Langevin term
\begin{equation}
    \delta_T J = \frac{\partial J}{\partial h}\delta_T h = \frac{\partial J}{\partial h} \int_0^{\tau_0} d\tau\, v\delta\dot v 
    = - \tau_0 \int_0^{\tau_0} d\tau\, v \delta i(\tau).
\end{equation}
Here, $\delta$ indicates a fluctuating Langevin contribution and $\delta_T$ that the quantity is integrated over a period $\tau_0$. We can then compute the variance
\begin{equation}
  \langle (\delta_T J)^2 \rangle = \tau_0^2 \int_0^{\tau_0} d\tau\, v^2 2\theta \frac{i_d(v)}{v}
  =2\theta \tau_0^2 \oint d\varphi\, i_d(v)
\end{equation}
Using that 
\begin{equation}
    \langle (\delta_T J)^2 \rangle = 2D\tau_0,
\end{equation}
we find the diffusion coefficient
\begin{equation}
  D = \frac{2\pi\theta}{\omega(J)} \oint d\varphi\, i_d(v) 
\end{equation}
of the action. 

Combining these results, we obtain the drift-diffusion equation
\begin{equation}
    \frac{\partial p}{\partial \tau} =   \frac{\partial}{\partial J}\left\{ \varepsilon_d(J) \left[ 1 +
    \frac{2\pi\theta}{\omega(J)}  
    \frac{\partial}{\partial J}
    \right]\right\} p.
\end{equation}
This generalizes the familiar result for the Ohmic case \cite{Haenggi1990} to general current-phase relations and non-Ohmic dissipative currents.  

Friction enters into the drift-diffusion equation only via $\varepsilon_d=\oint d\varphi\, i_d(v)$. Note that here, the integral is over one period of the trapped motion of the phase particle in the absence of friction.
The parameter $\varepsilon_d$ is the average energy dissipated per period of the trapped motion (in dimensionless units). Indeed, we have
\begin{equation}
    \varepsilon_d=\oint d\varphi\, i_d(v)
    = \frac{1}{I_c} \oint d\varphi\, I_d(V)
    = \frac{2e}{\hbar I_c} \oint dt\, V(t) I_d(V) 
    = \frac{E_d}{E_J},
\end{equation}
where $E_d$ is the dissipated energy per period in dimensional units.

Writing a differential equation for the distribution function $p(J)$ requires that $p(J)$ be smooth on the scale of the dissipated energy $\varepsilon_d$ per period. Provided we consider configurations sufficiently close to equilibrium, this is guaranteed as long as $ \varepsilon_d\ll \theta$. 

\subsection{Symmetry under sign change of bias current}

The drift-diffusion equation is invariant under sign changes of the bias current, provided that the untilted washboard potential $u_0(\varphi)$ is symmetric, $u_0(\varphi)=u_0(-\varphi)$.

To see this, we note that the phase-space trajectory $v=\pm v(\varphi)$ of energy $\varepsilon$ is given by
\begin{equation}
    v(\varphi) =\sqrt{2[\varepsilon  - u_0(\varphi)+i_b\varphi ]}.
\end{equation}
Without loss of generality, let's assume that the trapped motion is taking place in the minimum around $\varphi=0$ between the turning points $\varphi_L$ and $\varphi_R$, where $-\pi<\varphi_L<0$ and $0<\varphi_R<\pi$. The turning points solve
\begin{equation}
    \varepsilon=u_0(\varphi)-i_b\varphi.
\end{equation}
We note that 
\begin{equation}
    v(\varphi; i_b) = v(-\varphi;-i_b)
\end{equation}
and 
\begin{equation}
    \varphi_L(i_b)=-\varphi_R(-i_b)
    \quad ; \quad
    \varphi_R(i_b)=-\varphi_L(-i_b). 
\end{equation}

We first note that the action $J$ is invariant under sign changes of $i_b$. In fact, \begin{eqnarray}
   J(i_b)
   = 2\int_{\varphi_L(i_b)}^{\varphi_R(i_b)} d\varphi\, v(\varphi,i_b)
   =2\int_{-\varphi_R(i_b)}^{-\varphi_L(i_b)} d\varphi\, v(-\varphi,i_b)
   =2\int_{\varphi_L(-i_b)}^{\varphi_R(-i_b)} d\varphi\, v(\varphi,-i_b) = J(-i_b).
\end{eqnarray}
This implies that the relation $\varepsilon=\varepsilon(J)$ is symmetric as is the frequency $\omega(J)$.
Moreover, the disspation parameter $\varepsilon_d=\oint d\varphi\, i_d(v)$ is also independent of the sign of the bias current. We have
\begin{equation}
   \oint d\varphi\, i_d(v)
   = \int_{\varphi_L}^{\varphi_R} d\varphi\, i_d(v) + \int_{\varphi_R}^{\varphi_L} d\varphi\, i_d(-v)
   = \int_{\varphi_L}^{\varphi_R} d\varphi\, [i_d(v)-i_d(-v)],
\end{equation}
so that  
\begin{align}
    \varepsilon_d(i_b) =&\
   \int_{-\varphi_R(-i_b)}^{-\varphi_L(-i_b)} d\varphi\, [i_d(v(-\varphi;-i_b))-i_d(-v(-\varphi;-i_b))]
    \nonumber\\
    =&\ 
    \int_{\varphi_L(-i_b)}^{\varphi_R(-i_b)} d\varphi\, [i_d(v(\varphi;-i_b))-i_d(-v(\varphi;-i_b))]  
    = \varepsilon_d(-i_b).
\end{align}
These results imply that the drift-diffusion equation does not depend on the sign of $i_b$. 

\subsection{Switching rate in the weak-damping limit}

Computing the switching rate in the weak-damping limit is now a standard escape problem from a metastable well and requires one to solve the stationary drift-diffusion equation for $p(J)$ with the boundary condition that $p(J_B)=0$ \cite{Haenggi1990}. Here, $J_B$ is the action of the separatrix between trapped and running solutions in the absence of damping (red dashed contour in Fig.\ 1c). The boundary condition arises with the assumption that any trajectory that reaches the separatrix switches into a running solution. Due to the finite switching rate, the probability current
\begin{equation}
    j = -\varepsilon_d(J) \left[ 1 +
    \frac{2\pi\theta}{\omega(J)}  
    \frac{\partial}{\partial J}
    \right]p
\end{equation}
is nonzero. This equation for $p(J)$ is readily solved,
\begin{equation}
    p(J) =
    \int_J^{J_B} dJ'\,
    \exp\left\{\int_J^{J'} dJ''\, \frac{\omega(J'')}{2\pi\theta}\right\}
    \frac{\omega(J')}{2\pi\theta}\frac{j}{\varepsilon_d(J')}.
\end{equation}
The switching rate from the trapped to the running state (in dimensionless units) can be found as the ratio of the current $j$ out of the trapped state and the total probability to be in the trapped state,
\begin{equation}
    \gamma_\mathrm{sw} = \frac{j}{\int_0^{J_B} dJ\, p(J)}. 
\end{equation}
Note that the denominator is finite despite the logarithmic divergence of $p(J)$ at $J=0$ (due to $\varepsilon_d(J)\sim J$ at small $J$). Inserting the solution for $p(J)$, we find for the inverse rate
\begin{equation}
    (\gamma_\mathrm{sw})^{-1} =
    \int_0^{J_B} dJ
    \int_J^{J_B} dJ'\,  
    \exp\left\{\int_J^{J'} dJ''\,  \frac{\omega(J'')}{2\pi\theta}\right\}
    \frac{\omega(J')}{2\pi\theta}\frac{1}{\varepsilon_d(J')}.
\end{equation}
Using that $\omega(J)=2\pi \frac{dh(J)}{dJ}$ with the Hamiltonian $h(J)$ of the motion, the exponent integrates to 
\begin{equation}
    \int_J^{J'} dJ''\,  \frac{\omega(J'')}{2\pi\theta}
    = \frac{1}{\theta}[h(J')-h(J)],
\end{equation}
so that 
\begin{equation}
    (\gamma_\mathrm{sw})^{-1} =
    \int_0^{J_B} dJ\, 
    \exp\left\{-\frac{h(J)}{\theta}
    \right\}
    \int_J^{J_B} dJ' \, 
    \exp\left\{\frac{h(J')}{\theta}
    \right\}
    \frac{\omega(J')}{2\pi\theta}\frac{1}{\varepsilon_d(J')}.
\end{equation}
The $J'$ integral can be converted into an integral over $h$, which, in the limit of small temperatures, is dominated by the upper limit, 
\begin{equation}
    \int_J^{J_B} dJ' \, 
    \exp\left\{\frac{\varepsilon(J')}{\theta}
    \right\}
    \frac{\omega(J')}{2\pi\theta}\frac{1}{\varepsilon_d(J')}
    =\int_{h(J)}^{h(J_B)} dh' \,  \exp\left\{\frac{h'}{\theta}
    \right\}
    \frac{1}{\theta}\frac{1}{\varepsilon_d(J(h'))} \simeq
    \exp\left\{\frac{h(J_B)}{\theta}\right\}
    \frac{1}{\varepsilon_d(J_B)}.
\end{equation}
Note that this expression is independent of $J$, so that the $J$ integration now involves only 
\begin{equation}
    \int_0^{J_B} dJ\, 
    \exp\left\{-\frac{h(J)}{\theta}
    \right\}
    \simeq \frac{2\pi\theta}{\omega(J=0)}\exp\left\{-\frac{h(J=0)}{\theta}
    \right\}.
\end{equation}
Here, we used that the integral is dominated by the lower limit. Insertion into the expression for the switching rate
gives
\begin{equation}
    \gamma_\mathrm{sw}
    = \frac{\varepsilon_d(J_B) \omega_0}{2\pi \theta}
    \exp\left\{-\frac{\varepsilon_B}{\theta}\right\},
    \label{eq:gammatr}
\end{equation}
where $\varepsilon_B=h(J_B)-h(J=0)$ is the barrier height and $\omega_0=\omega(J=0)$. Reverting to dimensional parameters, this becomes
\begin{equation}
    \Gamma_\mathrm{sw}  = \frac{E_d(J_B) \Omega_0}{2\pi T}
    \exp\left\{-\frac{E_B}{T}\right\}.
\end{equation}
Note that $\Omega_0$ is the oscillation frequency about the minimum of the tilted washboard potential and is thus different from the plasma frequency $\Omega_p$.

\subsection{Switching current}
\label{sec:switchings}

When ramping up (or down) the bias current $i_b$ from zero (ramp rate $a>0$),
\begin{equation}
    i_b(\tau)=\pm a\tau,
    \label{eq:itau}
\end{equation}
the junction will eventually escape from the trapped state and abruptly switch into the running state. The bias levels at which this is happening are the switching currents $i_{\mathrm{sw},\pm}$. The probability $P_t$ for the system to be in the trapped state satisfies the rate equation
\begin{equation}
    \frac{dP_t}{d\tau} = -\gamma_\mathrm{sw}(\tau)P_t.
\end{equation}
The escape rate $\gamma_\mathrm{sw}(\tau)$ from the trapped to the running state depends on time, as the barrier height depends on the bias current. Solving for $P_t(\tau)$ with initial condition $P_t(\tau)=1$ gives
\begin{equation}
    P_t(\tau) = \exp\left\{-\int_0^\tau d\tau'\, 
    \gamma_\mathrm{sw}(\tau')
    \right\}.
\end{equation}
In view of Eq.\ (\ref{eq:itau}), we can replace the time argument by current,
\begin{equation}
    P_t(i_b) = \exp\left\{\mp\frac{1}{a}\int_0^{i_b} di_b'\, \gamma_\mathrm{sw}(i_b')
    \right\} =  \exp\left\{-\frac{1}{a}\int_0^{\abs{i_b}} di\, \gamma_\mathrm{sw}(\pm i)
    \right\}.
\end{equation}
This expression can be used to define the average switching currents $i_{\mathrm{sw},\pm} > 0$ through \begin{equation}
    P_t(i_b = \pm i_{\mathrm{sw},\pm})=\frac{1}{2},
\end{equation}
so that
\begin{equation}
    \int_0^{i_{\mathrm{sw},\pm} } di\,
    \gamma_\mathrm{sw}(\pm i) = a \ln 2.
\end{equation}
Inserting the escape rate $\gamma_\mathrm{sw}$ in Eq.\ (\ref{eq:gammatr}), this becomes
\begin{equation}
    a \ln 2
    = \int_0^{i_{\mathrm{sw},\pm}} di \,  \frac{\varepsilon_d(i) \omega_0(\pm i)}{2\pi \theta}
    \exp\left\{-\frac{\varepsilon_B(\pm i)}{\theta}
    \right\}.
\end{equation}
Due to the exponential factor, the integrand increases rapidly with increasing $i$ due to the decreasing barrier, so that the integral is dominated by bias currents $i \sim i_{\mathrm{sw},\pm}$. This allows us to approximate
\begin{equation}
    a \ln 2 \simeq \theta \bqty{  \abs{\frac{di_b}{d\varepsilon_B}} \gamma_\mathrm{sw} }_{i_b = \pm i_{\mathrm{sw},\pm}}.
\end{equation}
This yields
\begin{equation}
   \varepsilon_B(\pm i_{\mathrm{sw},\pm}) = \theta \ln \pqty{
   \frac{  \varepsilon_d(i_{\mathrm{sw},\pm}) \omega_0(\pm i_{\mathrm{sw},\pm})}{2\pi a \ln 2}\left|
    \frac{di_b}{d\varepsilon_B}
    \right|_{i_b=\pm i_{\mathrm{sw},\pm}}
   }.
\end{equation}
To make analytical progress, we approximate the bias-current dependence of the  barrier as
\begin{equation}
    \varepsilon_B(i_b) = \varepsilon_{B0} \left(1 \mp \frac{i_b}{i_{c,\pm}}\right)^{\mu_{\pm}},
\end{equation}
where $i_{c,\pm} > 0$ refers to the critical current for positive/negative bias and $\mu_\pm= \abs{\varphi_0^{\textrm{max}} - \varphi_0^{\textrm{min}} }_{\pm} i_{c,\pm}/\varepsilon_{B0}$, with $\abs{\varphi_0^{\textrm{max}} - \varphi_0^{\textrm{min}} }_{\pm} = \lim_{i_b \to 0^\pm} \abs{\varphi^{\textrm{max}} - \varphi^{\textrm{min}} }$.  This expression interpolates between  $\varepsilon_{B0}$ at zero bias and $\varepsilon_B=0$ at the critical current, and correctly reproduces the linear term in the small-$i_b$ expansion. It is  moreover a rather accurate approximation to set $i_{\mathrm{sw},\pm}\simeq 0$ under the logarithm (so that  $\omega_0\simeq 1$ in particular). We then find 
\begin{equation}
    i_{\mathrm{sw},\pm} \approx
    i_{c,\pm}\left\{1 - \left[
    \frac{\theta}{ \varepsilon_{B0} } \ln \left(
   \frac{ \varepsilon_d(J_{B0})}{2\pi a \ln 2 \abs{\varphi_0^{\textrm{max}} - \varphi_0^{\textrm{min}} }_{\pm}}
   \right)
    \right]^{1/\mu_\pm}\right\},
\end{equation}
where $J_{B0}$ is the action of the separatrix in the absence of damping and bias current. 
In dimensionful units, this is
\begin{equation}
    I_{\textrm{sw},\pm} = I_{c,\pm}\Bqty{1- \bqty{\frac{T}{E_B} \ln\pqty{  
    \frac{(2e/\hbar) \Omega_p E_d}{ 2\pi A \ln 2 \abs{\varphi_0^{\textrm{max}} - \varphi_0^{\textrm{min}} }_{\pm}   }  } }^{\frac{(2e/\hbar) E_B}{  \abs{\varphi_0^{\textrm{max}} - \varphi_0^{\textrm{min}} }_{\pm} I_{c,\pm} } }},
\end{equation}
where we defined the dimensionful barrier $E_B = E_J \varepsilon_{B0}$, dissipation $E_d = E_J \varepsilon_d(J_{B0})$, and ramp rate $A = dI_b/dt = (2eE_J/\hbar)\Omega_p a$. 

\section{Escape from the running state}

\subsection{Fokker-Planck equation in the weak-damping limit }

We now consider the escape rate from the running state. For weak friction, the retrapping current is small, so that we can restrict attention to the limit of small bias currents $i_b$ and thus of a weakly tilted washboard potential. In this case, we can consider the action variable for the untilted washboard potential $u_0(\varphi)$ as  slowly varying and satisfying a drift-diffusion equation. The phase variable is then equivalent to a pendulum in rotational motion, so that the action variable is given by 
\begin{equation}
    J = \int_0^{2\pi\, \mathrm{sgn} v} d\varphi\, v.
\end{equation}
The velocity is expressed in terms of the Hamiltonian 
\begin{equation}
    h_0 = \frac{v^2}{2} + u_0(\varphi)
\end{equation}
excluding the tilt. Note that the definition of the action differs from the definition in the context of the switching current. First, we now consider running solutions, so that the integral is over the full interval $[0,2\pi]$. Second, the action is evaluated for the deterministic motion in the absence of the current bias.   

There are now two contributions to the drift of the action $J$, one due to friction and a second due to the tilt of the washboard potential, 
\begin{equation}
    v_d = -\int_0^{2\pi\, \mathrm{sgn} v} d\varphi\, i_d(v) + 2\pi |i_b|.
\end{equation}
Let us assume for definiteness that the bias current is positive. Then, $v$ as well as $i_d(v)$ are positive, so that the friction force is negative. The opposite situation of negative bias currents follows by obvious changes of sign. Note again that unlike in the equation for the trapped state, the integral for the frictional contribution is over the interval $[0,2\pi]$.

The diffusion follows as in the equation for the trapped state,
\begin{equation}
  \langle (\delta_T J)^2 \rangle = \tau_0^2 \int_0^{\tau_0} d\tau\, v^2 2\theta \frac{i_d(v)}{v}
  =2\theta \tau_0^2 \int_0^{2\pi\,  \mathrm{sgn} v} d\varphi\, i_d(v).
\end{equation}
Here, the only change is in the integration range over $\varphi$, so that
\begin{equation}
  D = \frac{2\pi\theta}{\omega(J)} \int_0^{2\pi\, \mathrm{sgn} v} d\varphi\, i_d(v) .
\end{equation}
We can now combine these results into the drift-diffusion equation 
\begin{equation}
    \frac{\partial p}{\partial \tau} =   \frac{\partial}{\partial J} \left\{ \varepsilon_d(J)  -2\pi |i_b| + 
    \varepsilon_d(J)\frac{2\pi\theta}{\omega(J)} 
    \frac{\partial}{\partial J} \right\} p,
    \label{dd_run}
\end{equation}
where we defined
\begin{equation}
   \varepsilon_d(J) =   \int_0^{2\pi\,  \mathrm{sgn} v} d\varphi\, i_d(v).
   \label{edr}
\end{equation}
We will show in the next section that this Fokker-Planck equation reproduces the standard results for the retrapping rate of Ben-Jacob \textit{et al.} \cite{Benjacob1982}. Provided that the dissipative current is asymmetric, $i_d(v)\neq -i_d(-v)$, the dissipated energy in Eq.\ \eqref{edr} and the drift-diffusion equation (\ref{dd_run}) are now manifestly asymmetric under $i_b \to - i_b$.

By a similar argument as for the drift-diffusion equation for the trapped states, we find that the validity of Eq.\ (\ref{dd_run}) requires
\begin{equation}
    \frac{(\varepsilon_d - 2\pi i_b)^2}{\varepsilon_d} \ll \theta.
\end{equation}

\subsection{Retrapping rate}

In the absence of fluctuations, damping in the running state it is  
\begin{equation}
    \int_0^{2\pi\, \mathrm{sgn} v} d\varphi\, i_d(v)    = 2\pi |i_b|.
\end{equation}
Evaluating this for the (undamped, untilted) separatrix at action $J_B$ yields the 
retrapping current in the absence of  fluctuations as obtained in Eq.\ (\ref{eq:retrap_det}). (Note that $J_B$ as defined here coincides with $J_{B0}/2$ as defined in the calculation of the switching rate.) The retrapping current becomes small in the limit of weak damping, justifying our focus on small bias currents. 

We now solve the stationary drift-diffusion equation by using conservation of the probability current, 
\begin{equation}
    j = - \left\{ \varepsilon_d(J)   -2\pi |i_b| +
    \varepsilon_d(J)\frac{2\pi\theta}{\omega(J)}  
    \frac{\partial}{\partial J} \right\} p.
    \label{eq:jret}
\end{equation}
The stationary probability distribution with absorbing boundary condition at the separatrix, $p(J_B)=0$,
becomes
 \begin{equation}
     p(J) = -\int_{J_B}^J dJ''\, 
    \frac{j}{\varepsilon_d( J'')}\frac{\omega(J'')}{2\pi\theta}\exp\left\{\int_J^{J''} dJ'\, \frac{\omega(J')}{2\pi\theta}\left(1 - \frac{2\pi |i_b|}{\varepsilon_d(J')}\right)
    \right\}.
\end{equation}
Note that the probability current is negative (flow to smaller $J$), so that $p(J)$ is positive as it should.

The retrapping rate now follows as
\begin{equation}
    \gamma_\mathrm{re} = \frac{-j}{\int_{J_B}^\infty dJ\, p(J)},
\end{equation}
so that
\begin{equation}
    (\gamma_\mathrm{re})^{-1} = \int_{J_B}^\infty dJ\, 
    \int_{J_B}^J dJ'' \,  
    \frac{1}{\varepsilon_d( J'')}\frac{\omega(J'')}{2\pi\theta}\exp\left\{-\int^J_{J''} dJ'\, \frac{\omega(J')}{2\pi\theta}\left(1 - \frac{2\pi |i_b|}{\varepsilon_d( J')}\right)
    \right\}
\end{equation}
We will now simplify this expression in the low-temperature limit.

The evaluation can proceed with the following considerations:
(We present this derivation in some detail since it does not seem to be standard in the literature.)
\begin{itemize}
\item At low temperatures $\theta$, the integral over $J''$ is dominated by the minimum of 
\begin{equation}
    \int^J_{J''} dJ'\, \frac{\omega(J')}{2\pi\theta}\left(1 - \frac{2\pi |i_b|}{\varepsilon_d( J')}\right)
\end{equation}
in the integration range $J''\in [J_B,J]$. As a function of $J''$, this expression has an extremum at $\varepsilon_d( J_0)=2\pi i_b$. The action $J_0$ is just the stationary action in the limit of vanishing fluctuations, i.e., the action of the trajectory for which the energy gain due to the bias current and the frictional energy loss just compensate. We anticipate that even in the presence of weak fluctuations (low temperature), the distribution $p(J)$ is  sharply peaked around $J=J_0$, so that the $J$ integration is dominated by $J$ of the order $J_0$.
 
\item The integrand in the exponent is negative for $J''\ll J_0$ and positive for $J''\gg J_0$. Thus, as $J''$ increases from $J_B$, the integral in the exponent increases and the extremum is a maximum. (Note that $J''$ enters the exponent as the \textit{lower} integration limit.) Thus, the $J''$ integral is dominated by one of the integration limits, $J_B$ or $J$. In the first case, $J''\sim J_B$, the exponent is equal to 
\begin{equation}
    -\int^J_{J_B} dJ'\, \frac{\omega(J')}{2\pi\theta}\left(1 - \frac{2\pi |i_b|}{\varepsilon_d( J')}\right)
\end{equation}
to leading order. This grows with $J$ and is positive as long as $J<J_0$. In contrast, in the second case, $J''\sim J$, the exponent is equal to zero to leading order. The $J''$ integral is thus dominated by $J''\sim J_B$, implying also that the $J$ integral is dominated by $J\sim J_0$ as anticipated above. 

\item Expanding the exponent about $J''=J_B$ to linear order gives
\begin{equation}
-\int^J_{J''} dJ'\, \frac{\omega(J')}{2\pi\theta}\left(1 - \frac{2\pi |i_b|}{\varepsilon_d( J')}\right)
  \simeq 
    -\int^J_{J_B} dJ'\, \frac{\omega(J')}{2\pi\theta}\left(1 - \frac{2\pi |i_b|}{\varepsilon_d( J')}\right)
    +\frac{\omega(J_B)}{2\pi\theta}\left(1 - \frac{2\pi |i_b|}{\varepsilon_d( J_B)}\right)(J''-J_B).
\end{equation}
Thus, the $J''$ integration gives
\begin{equation}
    (\gamma_\mathrm{re})^{-1} = 
    \frac{1}{\varepsilon_d( J_B)}\frac{-1}{1-\frac{2\pi |i_b|}{\varepsilon_d( J_B)}}
    \int_{J_B}^\infty dJ\,  \exp\left\{-\int^J_{J_B} dJ'\, \frac{\omega(J')}{2\pi\theta}\left(1 - \frac{2\pi |i_b|}{\varepsilon_d( J')}\right)
    \right\}.
\end{equation}

\item The $J$ integral is now done by a standard saddle-point integration. Introducing the abbreviation 
\begin{equation}
    F(J) = \frac{\omega(J)}{2\pi\theta}\left(1 - \frac{2\pi |i_b|}{\varepsilon_d( J)}\right),
\end{equation}
we have
\begin{equation}
    (\gamma_\mathrm{re})^{-1} = 
    \frac{1}{\varepsilon_d( J_B)}\frac{-1}{1-\frac{2\pi |i_b|}{\varepsilon_d( J_B)}}
    \sqrt{\frac{2\pi}{F'(J_0)}}
    \exp\left\{-\int^{J_0}_{J_B} dJ'\, \frac{\omega(J')}{2\pi\theta}\left(1 - \frac{2\pi |i_b|}{\varepsilon_d( J')}\right)
    \right\}.
\end{equation}
Remember that $2\pi |i_b| = \varepsilon_d(J_0)$ and $2\pi i_{r0,+} = \varepsilon_d(J_B)$.
\end{itemize}

We finally bring this expression into a more compact form. First consider the exponent,
\begin{align}
    -\int^{J_0}_{J_B} dJ'\,\frac{\omega(J')}{2\pi\theta}\left(1 - \frac{2\pi |i_b|}{\varepsilon_d( J')}\right)
   =&\ -\int^{J_0}_{J_B} dJ'\,\frac{\omega(J')}{2\pi\theta}\frac{\varepsilon_d(J') - \varepsilon_d(J_0)} {\varepsilon_d( J')} 
    \nonumber\\
    \simeq&\
   - \frac{\omega(J_0)}{2\pi\theta}
   \frac{1} {\varepsilon_d( J_0)}\frac{1} {d\varepsilon_d( J_0)/dJ}
   \int^{\varepsilon_d(J_0)}_{\varepsilon_d(J_B)} d\varepsilon_d \,[\varepsilon_d - \varepsilon_d(J_0)]
    \nonumber\\
    \simeq&\
    \frac{\omega(J_0)}{2\pi\theta}
   \frac{1} {\varepsilon_d( J_0)}\frac{2\pi^2} {d\varepsilon_d( J_0)/dJ}
    [i_b \mp i_{r0,\pm}]^2.
\end{align}
Here, we used that in the weak-damping region, the retrapping rate is appreciable only in a narrow region near the retrapping current. Assuming that $|i_b|$ is not too close to the retrapping current $i_{\mathrm{r}0,\pm}$, we can employ a large-$\varepsilon$ expansion. Up to corrections of order $1/\varepsilon^2$, we find 
\begin{equation}
    J_0 = 2\pi \sqrt{2(\varepsilon_0-\overline{u}_0)},\  \omega(J_0) = \sqrt{2(\varepsilon_0-\overline{u}_0)},\ \varepsilon_d(J_0) = 2\pi i_d\pqty{\sqrt{2(\varepsilon_0-\overline{u}_0)}},\ \frac{d\varepsilon_d(J_0)}{dJ} = i^\prime_d\pqty{\sqrt{2(\varepsilon_0-\overline{u}_0)}}, 
\end{equation}
where $\varepsilon_0$ is related to the bias current through 
\begin{equation}
    i_b = i_d\pqty{\sqrt{2(\varepsilon_0-\overline{u}_0)}}.
\end{equation}
Here, $\overline{u}_0$ is the average value of $u_0(\varphi)$. Further introducing the average phase velocity $\overline{v}=\omega(J_0)$, we can express the exponent as 
\begin{equation}
    \frac{1}{2\theta}\frac{1}{\frac{i_d(\overline{v})}{\overline{v}} i_d^\prime(\overline{v})} [i_b \mp i_{r0,\pm}]^2.
\end{equation}

The preexponential factor in $\gamma_\mathrm{re}$ can be evaluated in the same manner and we finally find 
\begin{equation}
    \gamma_\mathrm{re}
    = \sqrt{\frac{i_d^\prime(\overline{v})}{i_d(\overline{v})/\overline{v}}\frac{[i_b\mp i_{r0,\pm}]^2}{2\pi\theta}}
    \exp\left\{-
    \frac{1}{2\theta}\frac{1}{\frac{i_d(\overline{v})}{\overline{v}} i_d^\prime(\overline{v})} [i_b \mp i_{r0,\pm}]^2\right\}
\end{equation}
for the retrapping rate with $i_b=i_d(\overline{v})$. One readily finds that for Ohmic dissipation and sinusoidal current-phase relation, this reduces to the retrapping rate as obtained by Ben-Jacob {\em et al}.\  \cite{Benjacob1982,Haenggi1990}. 

\subsection{Retrapping current}
The retrapping currents $i_{\mathrm{re},\pm}$ may be defined in a similar manner as the switching currents, see Sec.\ \ref{sec:switchings}. The bias current is now ramped down (up) from a sufficiently large starting value $\pm i_0$ with $i_0 \gg i_{\mathrm{re},\pm}$, i.e. 
\begin{equation}
    i_b(\tau) = \pm (i_0 - a \tau). 
\end{equation}
The average retrapping currents are defined as the bias level at which the probability for having retrapped reaches $1/2$. This gives 
\begin{align}
    \frac{1}{2} = \exp{\frac{1}{a}\int_{i_0}^{\pm i_{\mathrm{re},\pm}} di\ \gamma_\mathrm{re}(\pm i) },
\end{align}
Due to the exponential factor in $\gamma_{\mathrm{re}}$ the main contribution to the integral stems from $i_b \sim i_{\mathrm{re},\pm}$. This allows one to approximate
\begin{align}
    a \ln 2 \simeq&\ \frac{1}{\sqrt{2\pi\theta}}  \sqrt{\bqty{\frac{i_d^\prime(\overline{v})}{i_d(\overline{v})/\overline{v}}  }}_{i_b = \pm i_{\mathrm{re},\pm}}   \int_{i_{\mathrm{re},\pm}}^{\infty} di\, (i \mp i_{\mathrm{r}0,\pm}) \exp{-\bqty{\frac{1}{\frac{i_d(\overline{v})}{\overline{v}} i_d^\prime(\overline{v})} }_{i_b = \pm i_{\mathrm{re},\pm}} \frac{(i \mp i_{\mathrm{r}0,\pm})^2}{2\theta} } \\
    =&\ 
    \sqrt{\frac{\theta}{2\pi}\bqty{  \frac{i_d(\overline{v})}{\overline{v}}
    \pqty{ i_d^\prime(\overline{v})}^3 }  }_{i_b = \pm i_{\mathrm{re},\pm}}
    \exp{-\bqty{\frac{1}{\frac{i_d(\overline{v})}{\overline{v}} i_d^\prime(\overline{v})} }_{i_b = \pm i_{\mathrm{re},\pm}} \frac{(i_{\mathrm{re},\pm} - i_{\mathrm{r}0,\pm})^2}{2\theta} }. 
\end{align}
Setting $i_{\mathrm{re},\pm} \to i_{\mathrm{r}0,\pm}$ in the preexponential factor and solving for $i_{\mathrm{re},\pm}$ gives
\begin{equation}
    i_{\mathrm{re},\pm} = i_{\mathrm{r}0,\pm} + \Bqty{\theta  \frac{i_d(\overline{v})}{\overline{v}} i_d^\prime(\overline{v}) \ln \bqty{ \frac{\theta}{ 2\pi (a \ln 2)^2} 
     \frac{i_d(\overline{v})}{\overline{v}} \pqty{i_d^\prime(\overline{v})}^3
    } }^{1/2}_{i_b = \pm i_{\mathrm{r}0,\pm} }.
\end{equation}
In dimensionful units, this becomes 
\begin{equation}
    I_{\mathrm{re},\pm} = I_{\mathrm{r}0,\pm} + \Bqty{\frac{T}{C} \frac{I_d(\overline{V})}{\overline{V}} I_d^\prime(\overline{V}) \ln \bqty{ \frac{1}{ 2\pi (\ln 2)^2} \frac{T}{C^3 A^2}
     \frac{I_d(\overline{V})}{\overline{V}} \pqty{I_d^\prime(\overline{V})}^3
    } }^{1/2}_{I_b = \pm I_{\mathrm{r}0,\pm} },
\end{equation}
where $A = dI_b/dt$ is the dimensionful ramprate. 

\section{Yu-Shiba-Rusinov Josephson junction}\label{sec:ysr}
We consider a Josephson junction formed by a superconducting scanning tunneling microscope tip, labeled $L$, and a superconducting substrate in the presence of a magnetic impurity, labeled $R$. The magnetic impurity induces Yu-Shiba-Rusinov (YSR) bound states in the superconducting gap of the substrate. Importantly, the spectral weight for tunneling of electrons and holes into the YSR state is not equal, leading to asymmetric dissipative tunneling and hence a physical mechanism for asymmetric friction. 

\subsection{Junction Hamiltonian and YSR bound state}
We assume that tunneling of strength $\mathcal{T}$
is local at the site of the impurity. The Hamiltonian of the system is 
\begin{align}
    &H = \sum_{j = L/R} H_j + H_{\textrm{tun}} ,\\
    &H_{j} = \int d\mathbf{r}\ \Bqty\Big{ \sum_{\sigma\sigma'} \psi^{\dagger}_{j}(\mathbf{r}) \mathcal{H}_{j}  \psi_{j}(\mathbf{r})  + \Delta \bqty{  \psi^{\dagger}_{j,\uparrow}(\mathbf{r})\psi^{\dagger}_{j,\downarrow}(\mathbf{r})  + \textrm{h.c.}}  }, \\ 
    &H_{\textrm{tun}} = \mathcal{T}\sum_{\sigma}\bqty{ e^{i\varphi/2}  \psi^{\dagger}_{L,\sigma}(\mathbf{r}_0)  \psi_{R,\sigma}(\mathbf{r}_0) + \textrm{h.c.}},
\end{align}
with normal state single-particle Hamiltonians
\begin{equation}
    \mathcal{H}_L = \xi(-i\nabla),\ \mathcal{H}_R = \xi(-i\nabla) + (\mathcal{K} - \mathcal{J}\sigma_z)\delta(\mathbf{r}-\mathbf{r}_0).
\end{equation}
We assume that the superconductors are identical up to the presence of the magnetic impurity (characterized by exchange scattering $\mathcal{J}$ and potential scattering $\mathcal{K}$). We choose a gauge such that the pairing $\Delta$ is real in both the tip ($L$) and the substrate ($R$), such that the phase difference $\varphi$ enters the tunneling Hamiltonian. We define the local retarded Green functions of the tip and substrate as
\begin{align}
    G_{L}(E) =&\ \bra{\mathbf{r}_0}\Bqty{E_+ - \xi(-i\nabla)\tau_z - \Delta \tau_x}^{-1} \ket{\mathbf{r}_0} = -  \pi \nu_0 \frac{E_+ + \Delta \tau_x}{\sqrt{\Delta^2 - E_+^2}} , \\
    G_{R}(E) =&\  \Bqty{[G_{L}(E)]^{-1} + \mathcal{J}  - \mathcal{K} \tau_z  }^{-1} = - \pi \nu_0 \frac{  E_+ + \Delta \tau_x  + (\alpha + \beta \tau_z ) \sqrt{\Delta ^2-E_+^2}   }{(1-\alpha ^2+\beta ^2) \sqrt{\Delta ^2-E_+^2} - 2 \alpha  E_+},
\end{align}
where $\nu_0$ is the normal density of state at the Fermi level, $E_+ = E + i \eta$, with quasiparticle broadening$\eta$. The dimensionless exchange and potential scattering strengths are $\alpha = \pi\nu_0 \mathcal{J} > 0$ and $\beta  = \pi\nu_0 \mathcal{K}$.

The magnetic impurity induces a YSR bound state in the superconducting gap with energy
\begin{align}
    E_{\textrm{YSR}} = \Delta \frac{1-\alpha^2+\beta^2}{D} \quad , \quad  D = \sqrt{(1-\alpha^2+\beta^2)^2 + 4 \alpha^2} > 1
\end{align}
manifesting as a subgap pole in the substrate Green function. Importantly, the spectral weight differs for electron and hole parts as can be seen by expanding around $E_{\textrm{YSR}}$,
\begin{equation}
    G_{R}(E) \simeq \frac{2\pi\nu_0 \Delta}{E_+ - E_{\textrm{YSR}}} \begin{pmatrix} u^2 & uv \\ uv & v^2 \end{pmatrix};\ u,v = \frac{\sqrt{\alpha[1+(\alpha\pm\beta)^2]}}{D^3}.
\end{equation}

\subsection{RCSJ model for YSR tunnel junction}
We now discuss the phase dynamics of this YSR junction. In the tunneling limit, 
\begin{equation}
    \gamma \equiv (\pi\nu_0\mathcal{T})^2 \ll 1,
\end{equation}
the equation of motion for $\varphi$ is given by Eq.\ \eqref{eq:Langevin_dim}, with 
$ I_0(\varphi) = \frac{2e}{\hbar} E_J \sin \varphi$ and 
the quasiparticle tunneling current
\begin{align}\label{eq:quasiparticle_current}
    I_d(V) =&\ 
    \int \frac{dE}{2eR_{\textrm{tun}}} \bqty{n_F\pqty{E - \tfrac{eV}{2}} - n_F\pqty{E + \tfrac{eV}{2}}}    \bqty{ \mathcal{A}^e_L \pqty{E - \tfrac{eV}{2}}\mathcal{A}^e_R \pqty{E + \tfrac{eV}{2}} + \mathcal{A}^h_L \pqty{E + \tfrac{eV}{2}}\mathcal{A}^h_R \pqty{E - \tfrac{eV}{2}} },
\end{align}
in terms of the Fermi distribution function $n_F$ and the normal state tunneling resistance (w/o impurity) $R_{\textrm{tun}} = h/(8\gamma e^2)$. $\mathcal{A}^{e/h}_{L/R}$ is the (electron/hole) BCS density of states defined as
\begin{equation}
    \mathcal{A}^{e/h}_{L/R}(E) = -\frac{1}{\pi\nu_{0}} \Im \bqty{G^{e/h}_{L/R}(E)}.
\end{equation}
For $\alpha , \beta \neq 0$, electron and hole tunneling have different weight as $\mathcal{A}^{e}_{R} \neq \mathcal{A}^{h}_{R}$. At the same time, inversion symmetry of the junction is broken, $\mathcal{A}^{e/h}_{L} \neq \mathcal{A}^{e/h}_{R}$. As a consequence, $I_d(V) \neq -I_d(-V)$. 
Finally, the Josephson energy in the presence of the impurity is given by 
\begin{align}\label{eq:I_c_YSR}
    E_J = \frac{\gamma}{D} \Delta,
\end{align} 
This gives the plasma frequency as
\begin{equation}
    \Omega_p = \sqrt{\frac{\pi}{D}} \sqrt{\frac{\Delta/\hbar}{R_{\textrm{tun}}C}}.
\end{equation}
The parameters of the model are $\Delta,T,\eta,\Omega_p,\gamma,\alpha$ and $\beta$. We evaluate the quasiparticle current numerically and use that as an input for the Langevin dynamics as prescribed by the RCSJ model. 

\end{widetext}

\end{document}